\documentclass{pasj00}

\begin{document}
\SetRunningHead{S. Frey et al.}
{Identification of Potential Weak Target Radio Quasars for ASTRO-G In-Beam Phase-Referencing}
\Received{2008/08/05}
\Accepted{2008/10/07}

\title{Identification of Potential Weak Target Radio Quasars for ASTRO-G In-Beam Phase-Referencing}



\author{%
    S\'andor \textsc{Frey}\altaffilmark{1,2}
    Krisztina \'Eva \textsc{Gab\'anyi}\altaffilmark{3,2,1}
    and
    Yoshiharu \textsc{Asaki}\altaffilmark{3,4}}
\altaffiltext{1}{F\"OMI Satellite Geodetic Observatory,
P.O. Box 585, H-1592 Budapest, Hungary}
\altaffiltext{2}{MTA Research Group for Physical Geodesy and Geodynamics,
P.O. Box 91, H-1521 Budapest, Hungary}
\altaffiltext{3}{Institute of Space and Astronautical Science, Japan Aerospace Exploration Agency\\
3-1-1 Yoshinodai, Sagamihara, Kanagawa 229-8510, Japan}
\altaffiltext{4}{Department of Space and Astronautical Science, School of Physical Sciences,\\
The Graduate University of Advanced Studies, 3-1-1 Yoshinodai, Sagamihara, Kanagawa 229-8510, Japan}
    \email{frey@sgo.fomi.hu, gabanyik@vsop.isas.jaxa.jp, asaki@vsop.isas.jaxa.jp}


\maketitle

\begin{abstract}
We apply an efficient selection method to identify potential weak Very Long Baseline Interferometry (VLBI) target quasars simply using optical (SDSS) and low-resolution radio (FIRST) catalogue data. Our search is restricted to within $12\arcmin$ from known compact radio sources that are detectable as phase-reference calibrators for ASTRO-G at 8.4~GHz frequency. These calibrators have estimated correlated flux density $>20$~mJy on the longest ground--space VLBI baselines. The search radius corresponds to the primary beam size of the ASTRO-G antenna. We show that $\sim$20 quasars with at least mJy-level expected flux density can be pre-selected as potential in-beam phase-reference targets for ASTRO-G at 8.4~GHz frequency. Most of them have never been imaged with VLBI. The sample of these dominantly weak sources offers a good opportunity to study their radio structures with unprecedented angular resolution provided by Space VLBI. The method of in-beam phase-referencing is independent from the ability of the orbiting radio telescope to do rapid position-switching manoeuvres between the calibrators and the nearby reference sources, and less sensitive to the satellite orbit determination uncertainties.
\end{abstract}

\section{Introduction}

Phase-referencing is a way to increase the sensitivity of the Space Very Long Baseline Interferometry (SVLBI) observations that provide extremely high angular resolution due to the baselines exceeding the Earth diameter. Phase-reference imaging in ground-based VLBI is usually done in cycles of interleaving observations between a weak target source and a nearby strong reference source. Delay, delay-rate and phase solutions obtained for the phase-reference calibrator are interpolated and applied for the target source within the atmospheric coherence time, thus increasing the coherent integration time on the weak target source.   

Unlike the first dedicated SVLBI satellite HALCA \citep{hira00}, the next-generation satellite ASTRO-G will be capable of rapid attitude changing manoeuvres. This, and the accurate orbit determination will allow us to observe suitable nearby reference--target source pairs in the traditional ``nodding'' style \citep{asak07}. There is another, technically less demanding method which does not require rapid changes in the space antenna pointing if the reference--target separation is so small that both sources are within the primary beam of the 9.3-m ASTRO-G paraboloid antenna ($\sim12\arcmin$ at 8.4~GHz). In this scenario, the ground-based part of the SVLBI network performs the usual reference--target switching cycles, while the space antenna remains pointed to the same celestial point. (The diameters of the ground-based VLBI antennas are at least a factor of $\sim3$ larger, and thus their primary beam sizes are considerably smaller than that of the orbiting antenna.)

Even though the ASTRO-G satellite is being designed to be able to perform rapid position-switching manoeuvres between the calibrators and the reference sources, investigating the in-beam phase-referencing possibilities could still be beneficial. Such observations can be done even if the spacecraft orbit reconstruction error is as high as several meters \citep{guir01}. In-beam phase-referencing saves the on-board resources of the satellite and thus, in general, may help maintaining its observing capabilities for a longer period of time. Should the regular phase-referencing observations become difficult or impossible for some technical reason (e.g. the eventual failure of a control momentum gyro towards the end of the satellite lifetime), the in-beam method would still be possible to use.  

Successful in-beam phase-referencing experiments have already been conducted with HALCA which could not quickly change its antenna pointing \citep{pori00,bart00,porc00,guir01}. However, the use of in-beam phase-referencing is severely limited by the small number of sufficiently close source pairs known in the sky. Generally speaking, for any given target source of interest, it is very unlikely to find a suitable phase-reference calibrator within the primary beam of even the relatively small-diameter space antenna.

One may reverse the usual logic and select the phase-reference calibrator sources first. Then it becomes possible to look for potential weak target sources that are located so close to one of the reference sources that in-beam phase-referencing observations with the orbiting antenna are feasible. 
A recent experience with the Deep Extragalactic VLBI-Optical Survey \citep[DEVOS]{moso06,frey08} has shown that a large fraction (85\%) of the objects that are unresolved at arcsecond scales in the radio as well as  optically identified with quasars are detected with phase-referenced ground-based VLBI observations at 5~GHz. A similar conclusion was reached by \citet{bour08} who observed a sample of nearly 450 optically and radio-selected quasars with the European VLBI Network (EVN) at 2.3 and 8.4~GHz, in order to densify the radio reference frame. They report a VLBI detection rate of $\sim90$\%. 

Building upon these results, here we apply the DEVOS selection scheme and show that as many as $\sim$20 quasars with at least mJy-level expected flux density can be pre-selected as potential in-beam phase-reference targets for ASTRO-G. This prospective sample may be large enough for a comparison of sub-milliarcsecond radio structures of weaker sources with those of brighter ones. Such a sample, most of which have never been studied with VLBI, would certainly contain individually interesting radio quasars as well. The suitability of the candidate sources could be verified with ground-based VLBI observations prior to the launch of ASTRO-G.

\section{Sample selection}

Potential phase-reference sources are easily found in the most complete and up-to-date source catalogue maintained by the NASA Goddard Space Flight Center (GSFC) VLBI group. The version we used here is 2008a\_astro \citep{astro08} which incorporates the Very Long Baseline Array (VLBA) Calibrator Survey (VCS) sources \citep{beas02,foma03,petr05,petr06,kova07,petr08}. To pick up the compact and bright objects that are most likely to give good signal-to-noise ratio with SVLBI, we estimated the 8.4-GHz correlated flux densities on the the longest possible SVLBI baselines to ASTRO-G (25~000~km). This was done by fitting a single circular Gaussian brightness distribution model to the source visibility data. The Difmap program \citep{shep94} was used for obtaining the fitted model component total flux density $S_0$ and angular size $\theta$ (full width at half maximum, FWHM). The correlated flux density is
\begin{equation}
S_{\rm c} = S_0 \, \exp {\left( \frac{-(\pi \, \theta \, B)^2}{4 \, \ln{2}} \right)}
\end{equation}
where $B$ is the baseline length (25~000~km) expressed in the unit of the wavelength (3.6~cm) \citep{pear95}. Based on the 5-GHz VSOP Survey data, \citet{hori04} found that a ``typical'' compact radio-loud active galactic nucleus contains $\sim40$~\% of its total flux density in sub-mas compact components. Since we used ground-only (VLBA) data to infer the source structures, our fitted sizes are likely overestimated, and thus the correlated flux densities on the longest ground--space baselines are underestimated. 

The expected fringe detection limit on the SVLBI baselines to ASTRO-G was determined by estimating the baseline sensitivity
\begin{equation}
\sigma = \frac{1}{\eta} \sqrt{\frac{SEFD_1 SEFD_2}{2 \, \Delta\nu \, t}}
\end{equation}
where $\eta$ is the efficiency factor, $SEFD_1$ and $SEFD_2$ are the system equivalent flux densities of the space and ground antenna, respectively, $\Delta\nu$ is the bandwidth and $t$ is the coherent integration time \citep{walk95}.
Assuming $\eta=0.88$ (2-bit sampling), the 5-$\sigma$ fringe detection limit using $t=3$~min integration on a baseline between a 25-m VLBA antenna and ASTRO-G is 26~mJy. For ASTRO-G, we appied the nominal $SEFD_1=6100$~Jy and $\Delta\nu=256$~MHz \citep{tsub08}. The typical zenith system equivalent flux density at 8.4~GHz for a VLBA antenna is $SEFD_2=307$~Jy \citep{ulve08}.

We kept sources in the list of our potential phase-reference calibrators with $S_{\rm c}>20$~mJy. The fringe detection limit could actually be improved by up to a factor of $\sim4$ by using more sensitive ground radio telescopes available at present. More than 760 sources passed this filtering, at least at one of the observing epochs for which VLBA visibility data were available for model fitting. These are actually the potential calibrator sources available in the sky (mainly north of about $-45\degree$ declintion), that could be detected by ASTRO-G at 8.4~GHz within 3 min integration. Note that the flux density and angular size of the sources may vary over the time scale of a few years or shorter. 

The next step was to look for other objects within $12\arcmin$ of the positions of each potential reference source. This angular separation equals to the primary beam size (half-power beam width, HPBW) of the ASTRO-G antenna at 8.4~GHz. In other words, the reference sources and the targets -- if found at all -- lie within this beam. Note that for source separations nearly as large as the HPBW, the space antenna has to be pointed between the reference and target positions. Both sources are placed at around the half-power point which reduces the signal-to-noise ratio by about 30\%.

The search was performed in the Sloan Digital Sky Survey (SDSS) Data Release 6 (DR6) data base\footnote{\tt http://www.sdss.org/dr6} \citep{adel08}. The choice of an optical catalogue is justified by our experience with the DEVOS \citep{moso06,frey08}. It has been shown that 85\% of the SDSS optical {\it quasars} (i.e. extragalaxtic objects with stellar appearance) that coincide with an unresolved ($<5\arcsec$ fitted angular size) and ``strong'' ($>20$~mJy integral flux density) radio source in the Very Large Array (VLA) Faint Images of the Radio Sky at Twenty-centimeters (FIRST) Survey list\footnote{\tt http://sundog.stsci.edu} \citep{whit97} are detected with phase-referenced ground-based VLBI at 5~GHz. Therefore the cross-comparison of the SDSS and FIRST lists provides us with an efficient tool to pick up potential VLBI target sources at cm wavelengths, with at least mJy-level correlated flux densities. These sources have optical magnitudes readily available from SDSS.

\section{Results and discussion}

We found a total of 23 potential in-beam phase-reference target objects (quasars) which are unresolved {\it both} in the optical and in the radio with the VLA in FIRST. These are located around 20 different phase-reference calibrators.
The complete list of the phase-reference sources together with the corresponding SVLBI targets is given in Table~\ref{ref-target-sample}. We list the names and the accurate coordinates of each reference source, along with the estimates of correlated flux density $S_{\rm c}$ (last column; the maximum value is given where more epochs were available). In the next line(s), for the target source(s) we give FIRST integral flux density at 1.4~GHz and total flux density at 4.85~GHz from the GB6 catalogue \citep{greg96} for the sources in the northern hemisphere where available. These two flux density values together are useful to infer the radio spectral index of the target sources in the few-GHz range. Note that the GB6 detection limit is $\sim18$~mJy and therefore the spectral information is incomplete. We also list SDSS $r$ optical magnitude and spectroscopic redshift where available. For easy reference, the unique SDSS object identifiers are also provided in the last column of Table~\ref{ref-target-sample}.

The majority of the objects we found have never been studied with mas-resolution radio imaging. Considering the DEVOS detection rate at 5~GHz \citep{frey08}, we estimate that a fairly large sample, about 20 such quasar--reference source pairs could be successfully targeted with in-beam phase-referenced SVLBI observations at the somewhat higher frequency of 8.4~GHz. One of the selected reference--target pairs (J1041+523A and B) has already been studied with VSOP \citep{porc00}.

Obviously many more similar pairs should exist in the sky. However, the SDSS and FIRST catalogues have a limited sky coverage therefore our straightforward pre-selection method cannot be applied to identify them. Among the potential bright and compact phase-reference sources we picked up, about 240 objects (32\%) are found in the regions surveyed by FIRST. Extrapolating our success rate in finding nearby targets, another $\sim40$ reference sources outside the FIRST and SDSS coverage area chould have weak VLBI target quasars within $12\arcmin$ angular separation. 

The main advantage of our method is its ease. It applies simple selection criteria on currently available catalogue data. The key point is the identification of the unresolved FIRST sources with {\it optical} quasars. This almost guarantees that a mas-scale radio component is present in the source -- even though at a small fraction of the total flux density in some cases \citep{frey08}. 

One could in principle look for possible target radio sources by imaging the close vicinity of the calibrator sources with e.g. the VLA. Even archival data could be used, since the majority of our calibrators have certainly been observed at several epochs. However, with the VLA field of view at 8.4~GHz, it would require multiple pointings to image the whole area corresponding to the ASTRO-G primary beam. On the other hand, the angular resolution at 8.4~GHz ($0.24\arcsec-8.4\arcsec$, depending on the array configuration) does not guarantee that the weaker nearby sources are compact at the $\sim3$ orders of magnitude finer angular scales of SVLBI. Multi-frequency VLA observations would be needed to determine the radio spectral index of the objects found, and to select flat-spectrum sources that are most likely compact. Such a survey could bring potential target sources that are weaker than the FIRST detection limit ($\sim1$~mJy) but would require considerable data reduction efforts. 

Our DEVOS-based selection method could obviously be applied with any feasible calibrator--target separation limit as well, providing potential targets for conventional (nodding-style) phase-referencing observations involving ASTRO-G.

\section{Conclusions}

We provide a list of prospective radio sources to be targeted with in-beam phase-referencing with the SVLBI satellite ASTRO-G. Nearly all of the 23 sources found can be studied at sub-mas angular resolution with SVLBI, even if they are weaker than the fringe detection limit. The close vicinity of the suitable phase-reference calibrators (within $12\arcmin$) guarantees that the space antenna pointing can be kept fixed during the entire observing, while the ground-based VLBI antennas are regularly switching between the reference and target sources. This considerably simplifies the operation of the space radio telescope compared to the attitude-switching mode to be applied for ``traditional'' phase-referencing that involves more distant target--reference source pairs.
 
The sample selection presented here is purely technically driven. However, sources at $\sim1-10$~mJy flux density levels are still unexplored at the superior angular resolution offered by SVLBI. Astrophysical applications of observing this sample may include individual studies of source structures. The number of accessible weak target sources is quite large to allow some statistical investigations of their sub-mas radio structures and brightness tempearatures. This could be compared with the results obtained for the brightest radio sources in e.g. the 5-GHz VSOP Survey \citep[and references therein]{dods08}. The suitability of the selected sources for SVLBI imaging could be conveniently checked and verified by ground-based snaphshot VLBI experiments well before the launch of ASTRO-G. Pre-launch studies may include observations at multiple frequencies to identify the most promising target candidates.

The perspectives for in-beam phase-referencing at higher ASTRO-G frequencies (22 and 43~GHz) seem much less promising. This is partly due to the smaller number of suitable strong phase-reference calibrators. On the other hand, at higher observing frequencies the primary beam size of the space antenna ($4\farcm6$ and $2\farcm3$ HPBW at 22~GHz and 43~GHz, respectively) is proportionately smaller. For example, only 4 of our target--reference pairs listed in Table~\ref{ref-target-sample} would lie within the 22-GHz primary beam, even if they could qualify as suitable objects for phase-referenced SVLBI observations.

\par
\vspace{1pc}\par
We thank the anonymous referee for constructive suggestions. 
This research was supported by the Hungarian Scientific Research Fund (OTKA K72515) and the Hungarian Space Office (TP-314). K\'EG acknowledges a fellowship received from the Japan Society for Promotion of Science. This research has made use of the NASA/IPAC Extragalactic Database (NED) which is operated by the Jet Propulsion Laboratory, California Institute of Technology, under contract with NASA. 
Funding for the SDSS and SDSS-II has been provided by the Alfred P. Sloan Foundation, the Participating Institutions, the National Science Foundation, the U.S. Department of Energy, the National Aeronautics and Space Administration, the Japanese Monbukagakusho, the Max Planck Society, and the Higher Education Funding Council for England. The SDSS Web Site is http://www.sdss.org/. The SDSS is managed by the Astrophysical Research Consortium for the Participating Institutions.


\begin{table*}
  \caption{The list of phase-reference calibrator sources and the corresponding potential target sources within the primary beam of the ASTRO-G antenna at 8.4~GHz.}\label{ref-target-sample}
  \begin{center}
    \begin{tabular}{lllrrrrrr}
  \hline              
Source name & \multicolumn{2}{c}{Equatorial coordinates (J2000)} & $\delta$ & FIRST & GB6 & SDSS & $z$ & $S_{\rm c}$ (calibrator)\\
            & R.A. & Dec.                                        &          & int   & int & $r$  &     & or\\
  & h\hspace{3mm}m\hspace{3mm}s & \hspace{4mm}$\degree$\hspace{4mm}$\arcmin$\hspace{3mm}$\arcsec$ & $\arcmin$ & mJy & mJy &  &  &  SDSS objID (target)\\
  \hline              
J0039$-$0942 & 00 39 06.29166 & $-$09 42 46.8878 &      &       &     &       &      & 50 mJy\\
             & 00 38 44.77    & $-$09 34 17.9    & 10.0 &  44.0 &     & 22.52 &      & 587727226768064842\\
             & 00 39 39.78    & $-$09 49 32.4    & 10.7 &  41.1 &     & 20.27 &      & 587727179527487724\\

J0750+1823   & 07 50 00.32994 &   +18 23 11.4072 &      &       &     &       &      & 341 mJy\\
             & 07 49 22.31    &   +18 28 37.3    & 10.5 &  32.7 &     & 20.51 &      & 588016878285620011\\

J0918+0946   & 09 18 38.56849 &   +09 46 52.9006 &      &       &     &       &      & 54 mJy\\
             & 09 18 33.86    &   +09 46 04.6    &  1.4 &  52.5 &     & 21.30 &      & 588017676620005839\\

J0919+3324   & 09 19 08.78712 &   +33 24 41.9430 &      &       &     &       &      & 45 mJy\\
             & 09 19 35.53    &   +33 29 26.5    &  7.3 &  25.5 &     & 19.78 &      & 587735662619001099\\

J0921+6215   & 09 21 36.23107 &   +62 15 52.1804 &      &       &     &       &      & 116 mJy\\
             & 09 21 02.99    &   +62 15 07.8    &  3.9 &  41.5 &     & 19.68 &      & 587737825682850030\\

J1006+0510   & 10 06 37.61089 &   +05 09 53.9838 &      &       &     &       &      & 81 mJy\\
             & 10 06 55.80    &   +05 03 24.7    &  7.9 &  25.6 &     & 19.65 & 3.09 & 587728881412997267\\

J1041+523A   & 10 41 46.78164 &   +52 33 28.2313 &      &       &     &       &      & 105 mJy\\
             & 10 41 48.89    &   +52 33 55.6    &  0.6 & 106.8 &     & 18.11 & 2.30 & 587733080806654014\\

J1148+5254   & 11 48 56.56900 &   +52 54 25.3234 &      &       &     &       &      & 37 mJy\\
             & 11 48 16.09    &   +52 58 59.2    &  7.6 & 103.5 &  65 & 20.08 &      & 587732136457994454\\

J1153+0955   & 11 53 48.52737 &   +09 55 54.8891 &      &       &     &       &      & 24 mJy\\
             & 11 53 13.69    &   +09 48 22.6    & 11.4 &  22.1 &     & 21.86 &      & 587734892752405055\\

J1302+5748   & 13 02 52.46528 &   +57 48 37.6093 &      &       &     &       &      & 240 mJy\\
             & 13 03 10.68    &   +57 43 36.8    &  5.6 &  92.3 &  59 & 20.44 &      & 587735666912198784\\

J1308+3546   & 13 08 23.70915 &   +35 46 37.1640 &      &       &     &       &      & 256 mJy\\
             & 13 08 33.90    &   +35 51 08.9    &  5.0 &  23.6 &     & 15.60 &      & 587739096456953907\\

J1332+4722   & 13 32 45.24642 &   +47 22 22.6677 &      &       &     &       &      & 97 mJy\\
             & 13 33 25.06    &   +47 29 35.3    &  9.9 &  44.7 &  35 & 18.43 & 2.62 & 587732482218262542\\

J1353+6324   & 13 53 58.84901 &   +63 24 32.4454 &      &       &     &       &      & 21 mJy\\
             & 13 54 30.27    &   +63 32 28.9    &  8.7 &  20.3 &     & 20.90 &      & 587728919519691175\\

J1422+3223   & 14 22 30.37896 &   +32 23 10.4401 &      &       &     &       &      & 133 mJy\\
             & 14 22 21.33    &   +32 22 46.2    &  2.0 & 109.9 &     & 19.96 &      & 587739130808828033\\

J1430+4204   & 14 30 23.74163 &   +42 04 36.4911 &      &       &     &       &      & 72 mJy\\
             & 14 31 14.02    &   +42 05 52.9    &  9.4 &  29.3 &     & 20.85 &      & 588017115054997636\\

J1543+0452   & 15 43 33.92577 &   +04 52 19.3195 &      &       &     &       &      & 35 mJy\\
             & 15 42 46.37    &   +04 51 13.9    & 11.9 &  23.3 &     & 19.76 &      & 587730022798655756\\

J1549+0237   & 15 49 29.43685 &   +02 37 01.1634 &      &       &     &       &      & 426 mJy\\
             & 15 49 09.11    &   +02 38 21.2    &  5.3 &  74.9 &     & 21.64 &      & 587729158448153314\\
             & 15 50 07.07    &   +02 36 07.6    &  9.4 &  75.3 &     & 20.32 & 2.37 & 587729158448284044\\

J1603+1554   & 16 03 38.06238 &   +15 54 02.3590 &      &       &     &       &      & 98 mJy\\
             & 16 03 58.23    &   +15 47 36.3    &  8.1 &  45.5 &     & 20.47 &      & 587739827676054084\\

J1635+3808   & 16 35 15.49297 &   +38 08 04.5006 &      &       &     &       &      & 196 mJy\\
             & 16 35 41.54    &   +38 07 43.3    &  5.1 &  20.5 &     & 21.42 &      & 587733609628107409\\

J1646+4059   & 16 46 56.85869 &   +40 59 17.1721 &      &       &     &       &      & 27 mJy\\
             & 16 46 04.59    &   +41 03 12.2    & 10.6 & 101.5 &  28 & 20.15 &      & 588007005269918175\\
             & 16 46 29.99    &   +40 52 03.2    &  8.8 &  43.5 &  24 & 21.51 &      & 588007005270049620\\
      \hline
    \end{tabular}
  \end{center}
Notes for the reference sources: Col.~1 -- source name; Col.~2 -- J2000 right ascension (h m s);
Col.~3 -- J2000 declination ($\degree$ $\arcmin$ $\arcsec$); coordinates are taken from \citet{astro08};
Col.~9 -- best estimated 8.4-GHz correlated flux density on 25~000~km ground--space VLBI baselines\\
Notes for the target sources: Col.~2 -- J2000 right ascension (h m s); Col.~3 -- J2000 declination ($\degree$ $\arcmin$ $\arcsec$); 
coordinates are taken from the SDSS DR6; Col.~4 -- reference--target angular separation ($\arcmin$); 
Col.~5 -- FIRST integral flux density at 1.4~GHz; Col.~6 -- GB6 total flux density at 4.85~GHz;
Col.~7 -- SDSS $r$ magnitude; Col.~8 -- SDSS spectroscopic redshift; Col.~9 -- SDSS object identifier\\
\end{table*}



\begin{thebibliography}{}{}

\bibitem[Adelman-McCarthy et al.(2008)]{adel08}
Adelman-McCarthy, J.K., Ag\"ueros, M.A., Allam, S.S., et al. 2008, \apjs, 175, 297 

\bibitem[Asaki et al.(2007)]{asak07}
Asaki, Y., Sudou, H., Kono, Y., et al. 2007, \pasj, 59, 397 

\bibitem[Bartel \& Bietenholz(2000)]{bart00}
Bartel, N., \& Bietenholz, M.F. 2000, in Astrophysical Phenomena Revealed by Space VLBI, ed. 
H. Hirabayashi, P.G. Edwards \& D.W. Murphy (Sagamihara: ISAS), 17 

\bibitem[Beasley et al.(2002)]{beas02}
Beasley, A.J., Gordon, D., Peck, A.B., et al. 2002, \apjs, 141, 13

\bibitem[Bourda et al.(2008)]{bour08}
Bourda, G., Charlot, P., Porcas, R., \& Garrington, S. 2008, in Proc. 5th IVS General Meeting, in press (arXiv:0804.3859)

\bibitem[Dodson et al.(2008)]{dods08}
Dodson, R., Fomalont, E.B., Wiik, K., et al. 2008, \apjs, 175, 314 

\bibitem[Fomalont et al.(2003)]{foma03}
Fomalont, E.B., Petrov, L., MacMillan, D.S. et al. 2003, \aj, 126, 2562

\bibitem[Frey et al.(2008)]{frey08}
Frey, S., Gurvits, L.I., Paragi, Z., et al. 2008, \aap, 477, 781 

\bibitem[Gregory et al.(1996)]{greg96}
Gregory, P.C., Scott, W.K., Douglas, K. \& Condon J.J. 1996, \apjs, 103, 427

\bibitem[Guirado et al.(2001)]{guir01}
Guirado, J.C., Ros, E., Jones, D.L., et al. 2001, \aap, 371, 766 

\bibitem[Hirabayashi et al.(2000)]{hira00}
Hirabayashi, H., Hirosawa, H., Kobayashi, H., et al. 2000, \pasj, 52, 955 

\bibitem[Horiuchi et al.(2004)]{hori04}
Horiuchi, S., Fomalont, E.B., Scott, W.K., et al. 2004, \apj, 616, 110 

\bibitem[Kovalev et al.(2007)]{kova07}
Kovalev, Y.Y., Petrov, L., Fomalont, E.B., \& Gordon, D. 2007, \aj, 133, 1236

\bibitem[Mosoni et al.(2006)]{moso06}
Mosoni, L., Frey, S., Gurvits, L.I., et al. 2006, \aap, 445, 413 

\bibitem[Pearson(1995)]{pear95}
Pearson, T.J. 1995, in ASP Conf. Ser. 82, Very Long Baseline Interferometry and the VLBA, ed. J.A. Zensus, P.J. Diamond, \& P.J. Napier (San Francisco: ASP), 267

\bibitem[Petrov(2008)]{astro08}
Petrov, L. 2008, VLBI global solution 2008a\_astro,\\
http://vlbi.gsfc.nasa.gov/solutions/2008a\_astro

\bibitem[Petrov et al.(2005)]{petr05}
Petrov, L., Kovalev, Y.Y., Fomalont, E.B., \& Gordon, D. 2005, \aj, 129, 1163

\bibitem[Petrov et al.(2006)]{petr06}
Petrov, L., Kovalev, Y.Y., Fomalont, E.B., \& Gordon, D. 2006, \aj, 131, 1872

\bibitem[Petrov et al.(2008)]{petr08}
Petrov, L., Kovalev, Y.Y., Fomalont, E.B., \& Gordon, D. 2008, \aj, 136, 580

\bibitem[Porcas \& Rioja(2000)]{pori00}
Porcas, R.W., \& Rioja, M.J. 2000, Adv. Space Res., 26, 673

\bibitem[Porcas et al.(2000)]{porc00}
Porcas, R.W., Rioja, M.J., Machalski, J., \& Hirabayashi H. 2000, in Astrophysical Phenomena Revealed by Space VLBI, ed. 
H. Hirabayashi, P.G. Edwards \& D.W. Murphy (Sagamihara: ISAS), 245 

\bibitem[Shepherd et al.(1994)]{shep94}
Shepherd, M.C., Pearson, T.J., \& Taylor, G.B. 1994, \baas, 26, 987

\bibitem[Tsuboi(2008)]{tsub08}
Tsuboi, M. 2008, Proc. SPIE, 7010, 701024

\bibitem[Ulvestad(2008)]{ulve08}
Ulvestad, J.S. (ed.) 2008, Very Long Baseline Array Observational Status Summary,\\
http://www.vlba.nrao.edu/astro/obstatus/current/ 

\bibitem[Walker(1995)]{walk95}
Walker, R.C. 1995, in ASP Conf. Ser. 82, Very Long Baseline Interferometry and the VLBA, ed. J.A. Zensus, P.J. Diamond, \& P.J. Napier (San Francisco: ASP), 133

\bibitem[White et al.(1997)]{whit97}
White, R.L., Becker, R.H., Helfand, D.J., \& Gregg, M.D. 1997, \apj, 475, 479

\end{thebibliography}
\end{document}